\documentclass[aps,prc,twocolumn,amsmath,amssymb,floatfix,superscriptaddress]
{revtex4-1}
\usepackage{silence} 
\usepackage{mathptmx}
\usepackage[T1]{fontenc}
\usepackage{CJK}
\usepackage{graphicx}
\usepackage{bm}
\usepackage{dcolumn}
\usepackage{multirow}
\usepackage{xcolor}
\usepackage[
breaklinks,pdfstartview=FitH,CJKbookmarks=true,
bookmarksnumbered=true,bookmarksopen=true,pdfborder={0 0 1},
colorlinks=true,linkcolor=blue,urlcolor=blue,anchorcolor=blue,citecolor=blue
]{hyperref}

\def\be{\begin{equation}} \def\ee{\end{equation}}
\def\bal#1\eal{\begin{align}#1\end{align}}
\def\bse#1\ese{\begin{subequations}#1\end{subequations}}

\def\ra{\rightarrow}
\long\def\hj#1{\color{red}#1\color{black}}
\long\def\OFF#1{}

\def\al{\alpha}
\def\eps{\varepsilon}
\def\la{\Lambda}
\def\rv{\bm{r}} \def\Jv{\bm{J}} \def\Wv{\bm{W}} \def\Nv{\bm{\nabla}}
\def\ron{\rho_N} 
\def\rol{\rho_\la}  
\def\las{\la_{s}} \def\lap{\la_{p}}
\def\bet{\beta_2}
\def\betn{\beta_2^{(N)}}

\def\fm{\,\text{fm}}
\def\fmq{\,\text{fm}^{-3}}
\def\mev{\;\text{MeV}}
\def\mfmq{\,\text{MeV}\,\text{fm}^{-3}}

\def\bl{B_\la}
\def\dbl{\delta B_\la}
\def\f16{^{16}\text{F}}
\def\o15{^{15}\text{O}}
\def\fls{^{17}_{\la s}\text{F}}
\def\flp{^{17}_{\la p}\text{F}}
\def\ols{^{16}_{\la s}\text{O}}
\def\olp{^{16}_{\la p}\text{O}}

\begin{document}

\begin{CJK*}{UTF8}{gbsn}

\title{Proton drip line of deformed hypernuclei}

\author{Yi-Xiu Liu (刘益秀)}
\affiliation{
	Department of Physics, East China Normal University, Shanghai 200241, China}
\author{Huai-Tong Xue (薛怀通)}\email{52204700001@stu.ecnu.edu.cn}
\affiliation{
	College of Physics and Electronic Engineering, Nanyang Normal University, Nanyang 473061, China}
\author{Q. B. Chen (陈启博)}
\author{Xian-Rong Zhou (周先荣)} \email{xrzhou@phy.ecnu.edu.cn}
\affiliation{
	Department of Physics, East China Normal University, Shanghai 200241, China}

\author{H.-J. Schulze}
\affiliation{
	INFN Sezione di Catania, Dipartimento di Fisica,
	Universit\'a di Catania, Via Santa Sofia 64, 95123 Catania, Italy}

\date{\today}

\begin{abstract}
The proton drip line of (hyper)nuclei is examined
within the framework of the deformed Skyrme-Hartree-Fock approach
by adjusting the nuclear force parameters to exactly reproduce
the core binding energies.
The impact of adding a $\la$ hyperon in a $s$ or $p$ state is studied,
and it is found that in some cases the deformation effect facilitates
the extension of the drip line by an added $p$-state hyperon.
However, no extension of the drip line is found for $s$-state hypernuclei.
\end{abstract}

\maketitle
\end{CJK*}

\section{Introduction}
\label{s:intro}

A rigorous test of a global nuclear theory is its ability
to predict the number of bound nuclides.
The boundaries of the nuclear chart are defined by the so-called drip lines,
beyond which no additional neutron or proton can be added to the nucleus.
Except for lighter systems,
the neutron drip line extends to isotopes far beyond
what is experimentally accessible,
whereas the proton drip line is much closer to stability,
due to the Coulomb-repulsion effects that increase with atomic number $Z$.

Proton- and neutron-rich nuclei close to their drip lines
have been intensively studied both experimentally
\cite{Symons79,Westfall79,Ozawa01,Jonson04,Kwan12,Bing22}
and theoretically
\cite{Kim22,Tang23,Lykiardopoulou23}.
In addition,
exotic phenomena often occur in those nuclei,
and therefore their investigation
has the potential to advance the field of exotic nuclear physics.
The study of exotic nuclei has thus attracted worldwide attention
\cite{Tanihata95,Mueller01}.

In nuclei with added strangeness,
strange particles might also affect the drip lines.
Previous studies have shown that
the weakly-bound nuclear cores of drip line nuclei
might be stabilized by the addition of further strongly-bound particles,
and the drip line might even be extended by added hyperons
\cite{Vretenar98,Tretyakova99,Zhou08,Samanta08,Gal13b,Hiyama15}
or kaons \cite{Guo22b}.

At present, there is considerable research on the properties of hypernuclei
with stable cores,
leading to relatively clear understanding of $\la$ hyperon-nucleon interactions.
However, studies on hypernuclei with unstable cores remain limited
\cite{YFChen22,Xue23,Nemura02,Hiyama00,Isaka11,Mei18,Cui17}.
Thus, our knowledge of the hypernuclear interactions
can be substantially improved
by investigating the hypernuclear chart beyond the stability valley.
Dalitz and Levi Setti \cite{Dalitz63} made the first theoretical treatment of
some neutron-rich $\la$ hypernuclei as early as 1963,
but the topic received little attention for many years.
Majling started a thorough investigation on
neutron-rich $\la$ hypernuclei in the 1990s \cite{Majling95},
and recent theoretical calculations demonstrate the
possible extension of the neutron drip line in $\la$ hypernuclei
as well as their halo structure
\cite{Vretenar98,Tretyakova99,Zhou08,Samanta08,Gal13b,Hiyama15,
Xue22,Sidorov22,Myo23}.

Experimentally,
neutron-rich hypernuclei ($^{6,8}_{\ \ \la}$He)
with unstable cores were indeed seen in early emulsion studies \cite{Davis86}.
At KEK, the first productive experiment dedicated to the
neutron-rich hypernucleus $^{10}_{\ \la}$Li
generated by the $^{10}$B$(\pi^-,K^+)$ process
was carried out \cite{Saha05},
but statistical error prevented the determination of its binding energy.
It was later discovered in other nuclei,
including $^6$He, $^{17,19}$B, $^{15,19,22}$C, and $^{29}$F \cite{Bagchi20}.
Regarding the most extreme case of the $^6_\la$H hypernucleus,
the FINUDA collaboration reported three candidate events
from the $^6$Li$(K^-,\pi^+)$ reaction \cite{Agnello12},
but the J-PARC E10 experiment \cite{Sugimura14}
failed to obtain a signal for the
$^6$Li$(\pi^-,K^+)^6_\la$H reaction.
In conclusion, there is still a dearth of experimental information
about neutron-rich $\la$ hypernuclei at this time.

Proton-rich hypernuclei have so far attracted even less focus,
although it is intriguing
that proton-rich isotopes are lighter than neutron-rich ones
and closer to the stability valley,
and can thus form more easily.
In emulsions, the hypernucleus $^7_\la$Be
with an unstable core $^6$Be has been seen \cite{Davis86}.
There are no current plans or particular experiments
aimed at proton-rich hypernuclei,
because they cannot be created
by the typical $(K^-,\pi^+)$, $(\pi^-,K^+)$, and $(e,e'K)$ reactions.
However, a novel avenue for the investigation of hypernuclei
is provided by heavy ion collisions.
Hypernuclear research was listed as
one of the goals of NICA, FAIR, and HIAF
\cite{Yang13,Botvina16,Rappold16,Feng20,Saito21,Saito21b,HIAF}.
The groundbreaking research in Dubna \cite{Avramenko92}
established the viability of medium-energy heavy ion collisions
for the production of hypernuclei,
and the HYPHI collaboration \cite{Saito16} verified these findings.
In principle, heavy ion experiments can produce hypernuclei of any composition,
including proton- and neutron-rich and even multi-$\la$ hypernuclei
\cite{Andronic11}.

Theoretically, proton-rich hypernuclei have recently been investigated
\cite{Lanskoy22,Kornilova23}.
While the known hypernucleus $^7_\la$Be 
\cite{Davis86,HyperChart}
represents the extension of the
Be-core drip line ($^6$Be is unbound),
in those works $^9_\la$C was claimed as extension of the C-core drip line,
but $^8_\la$B and $^{12}_{\ \la}$N were excluded from such roles.
Furthermore, the beyond-dripline core nuclei
$^{16}$F, $^{19}$Na, $^{16}$Ne, and $^{19}$Mg
were identified as possible candidates for an extension
of the proton drip line by an added $\la$.
However, definite predictions could not be made,
due to the very delicate dependence of the relevant
small proton separation energies on both the $NN$ and $N \la$ interactions.

In this article we strive to improve the accuracy of the theoretical
predictions regarding these nuclei.
In particular, we adjust the strength of the $NN$ interaction
in order to exactly reproduce the (small negative) proton separation energies
of the nuclear cores
and then study in detail the addition of a $\la$ in $s$ or $p$ states
to the proton-rich nucleus.
We use the deformed Skyrme-Hartree-Fock (DSHF) model
employing an accurate $\la N$ force
to compute the (hyper)nuclei
\cite{Zhou07,Zhou08,Schulze10,
Jin19,Jin20,Guo21,Guo22,Guo22b,Xue22,CFChen22,YFChen22,Liu23,Xue23,Xue24}
and pay close attention to how the system changes
when a $\la$ hyperon is introduced,
in particular regarding the effects of core deformation.

The paper is organized as follows.
In Sec.~\ref{s:form}, we briefly review the DSHF density-functional
approach to $\la$ hypernuclei.
In Sec.~\ref{s:res} we present the results regarding possible extension
of the core drip line for various hypernuclei.
The conclusions are given in Sec.~\ref{s:conc}.

\section{Formalism}
\label{s:form}

The SHF mean field method is
a powerful theoretical density functional method,
which can be applied comprehensively
from light nuclei to heavy nuclei.
In the mean-field theory,
the binding energy of the $\la$ hypernucleus is given by
\cite{Vautherin72,Vautherin73,Bender03,Stone07,Erler11,Schulze14,Rayet76,Rayet81}
\be
 E = \int d^3\rv\; \eps(\rv) \:,
\label{e:Energy}
\ee
where the energy-density functional is
\be
 \eps = \eps_N[\rho_n,\rho_p,\tau_n,\tau_p,\Jv_n,\Jv_p] +
 \eps_\la[\rho_N,\rol,\tau_\la,\Jv_N,\Jv_\la] \:,
\ee
with $\eps_N$ and $\eps_\la$ as contributions from
$NN$ and $\la N$ interactions, respectively.
For the nucleonic functional $\eps_N$,
we use the standard Skyrme forces
SLy4 \cite{Chabanat98} or
SkI4 \cite{Reinhard95}.
The one-body density $\rho_q$,
kinetic density $\tau_q$,
and spin-orbit (s.o.) current $\Jv_q$ read
\be
 \Big[ \rho_q,\; \tau_q,\; \Jv_q \Big] =
 \sum_{k=1}^{N_q} {n_q^k} \Big[
 |\phi_q^k|^2 ,\;
 |\Nv\phi_q^k|^2 ,\;
 {\phi_q^k}^* (\Nv\phi_q^k \times \bm{\sigma})/i
 \Big] \:,
\label{e:rho}
\ee
where $\phi_q^k$ $(k=1,\cdots,N_q)$
are the s.p.~wave functions
of the $k$-th occupied states for the different particles $q=n,p,\la$.

The occupation probabilities $n_q^k$ are calculated by taking into account
pairing within a BCS approximation for nucleons only.
The pairing interaction between nucleons is taken as a density-dependent
$\delta$ force \cite{Tajima93},
\be
\label{e:pair}
 V_q\left(\rv_1, \rv_2\right) = V_q'
 \left[ 1-\frac{\ron((\rv_1+\rv_2)/2)}{0.16\;\fmq} \right]
 \delta\left(\rv_1-\rv_2\right) \:,
\ee
where pairing strengths
$V'_p = V'_n = -410\;\text{MeVfm}^3$ are used
\cite{Zhou07,Suzuki03}.
A smooth energy cutoff is employed in the BCS calculations and
in the case of an odd nucleon number
the orbit occupied by the unpaired nucleon is blocked as described
in Ref.~\cite{Ring04}.

Through the variation of the binding energy Eq.~(\ref{e:Energy}) one derives
the SHF Schr\"odinger equation for both nucleons and hyperons,
\be
 \Big[ - \Nv \cdot \frac{1}{2m_q^*(\rv)} \Nv
 + V_q(\rv) -i\Wv_q(\rv) \cdot (\Nv\times\bm{\sigma})
 \Big] \phi_q^k(\rv) = e_q^k \phi_q^k(\rv) \:,
\label{e:se}
\ee
where $V_q(\rv)$ is the central part of the mean field depending
on the densities,
while $\Wv_q(\rv)$ is the s.o.~interaction part
\cite{Vautherin73,Bender99}.

For the Skyrme-type SLL4 $\la N$ interaction
that we will use in this work,
$\eps_\la$ is given as
\cite{Rayet81,Yamamoto88,Yamamoto10,Schulze14,Schulze19}
\bal
 \eps_\la =& \frac{\tau_\la}{2m_\la} + a_0 \rol\ron
 + a_3 \rol\ron^2
\notag\\&
 + a_1 \left(\rol \tau_N + \ron \tau_\la \right)
 - a_2 \left(\rol \Delta\ron + \ron \Delta\rol \right)/2
\notag\\&
 - a_4 \left(\rol\Nv\cdot\Jv_N + \ron\Nv\cdot\Jv_\la \right)
\:.
\label{e:ef}
\eal
\OFF{
where the last term is the s.o.~part,
Two alternative parametrizations of nonlinear effects are indicated,
i.e., the first one $a_3$ derived from a $G$-matrix
\cite{Millener88,Lanskoy97,Guleria12}
and the second one $a_3'$ from a $\la NN$ contact force
\cite{Rayet76,Rayet81}.
The SLL4 
LY1 \cite{Lanskoy97} 
$\la N$ force used in this work employs the first choice.
and the YBZ1 force \cite{Yamamoto88} the second one.
}
For convenience we repeat here the parameters \cite{Schulze19}:
$a_{0,1,2,3}=[ -322.0, 15.75, 19.63, 715.0 ]$
(in appropriate units for $\rho$ given in $\fmq$ and $\eps$ in $\mfmq$),
whereas the small s.o.~parameter $a_4$ \cite{Ajimura01,Motoba08,Xue23}
is neglected in this work.
We remind that these optimal force parameters were obtained
by fitting the current set of hypernuclear data
in combination with the SLy4 force,
so the predictions with the SLy4 force
may be considered slightly more accurate,
although the difference with the SkI4 results is usually very small.

\begin{table}[t]
\caption{
Experimental binding energy and single- and double-proton removal energies
(in MeV)
\cite{NuDat,Wang21}
of several light nuclei just beyond the proton drip line,
i.e., at least one of $S_p$ or $S_{pp}$ is negative.
Bold values indicate the critical decay.
}
\renewcommand\arraystretch{1.2}
\setlength{\tabcolsep}{1.5pt}
\def\mc#1{\multicolumn{1}{c}{$#1$}}
\begin{ruledtabular}
\begin{tabular}{ l r r >$r<$ >$r<$ }
         &\mc{Z}&\mc{-E} &\mc{S_p}   &\mc{S_{pp}} \\ 
\hline
$^{6}$Be   &  4 &  26.92 &     0.59  & \bm{-1.37} \\ 
$^{7}$B    &  5 &  24.91 &\bm{-2.01} &     -1.42  \\ 
$^{8}$C    &  6 &  24.81 &    -0.10  & \bm{-2.11} \\ 
$^{11}$N   &  7 &  58.94 &\bm{-1.38} &      2.63  \\ 
$^{12}$O   &  8 &  58.57 &    -0.36  & \bm{-1.74} \\ 
$^{16}$F   &  9 & 111.42 &\bm{-0.53} &      6.77  \\ 
$^{16}$Ne  & 10 &  97.33 &    -0.13  & \bm{-1.40} \\ 
$^{19}$Na  & 11 & 131.82 &\bm{-0.32} &      3.60  \\ 
$^{19}$Mg  & 12 & 112.12 &     0.49  & \bm{-0.76} \\ 
$^{21}$Al  & 13 & 132.24 &    -2.32  &      0.42  \\ 
$^{22}$Si  & 14 & 132.97 &     0.74  &     -1.58  \\ 
$^{25}$P   & 15 & 169.85 &    -2.16  &      1.14  \\ 
$^{26}$S   & 16 & 169.65 &    -0.20  &     -2.36  \\ 
$^{30}$Cl  & 17 & 224.17 &\bm{-0.48} &      2.76  \\ 
\end{tabular}
\end{ruledtabular}
\label{t:sp}
\end{table}

\begin{table*}[t]
\caption{
Properties of $^{16}$F and $^{15}$O and their hypernuclei
in the DSHF approach
(deformed and spherical results)
with original ($s_{t2}=s_{b4}=1$) and adjusted
SLy4 and SkI4 $NN$ forces:
quadrupole deformation $\betn$,
binding energy $-E$,
single-proton removal energy $S_p$,
$\la$ removal energy $B_\la$,
and their difference $\dbl$, Eq.~(\ref{e:dbl}).
All energies are in MeV.
}
\renewcommand\arraystretch{1.3}
\setlength{\tabcolsep}{1.5pt}
\def\mc#1{\multicolumn{2}{c}{#1}}
\def\md#1{\multicolumn{2}{c|}{#1}}
\begin{ruledtabular}
\begin{tabular}{ll|rrrr|rrrr}
&               & \multicolumn{4}{c|}{SLy4} & \multicolumn{4}{c}{SkI4}   \\
&               & def. & sph. & def. & sph. & def. & sph. & def. & sph.  \\
\hline
\md{$s_{t2}$}   & \mc{0.965}  & \md{1.000}  & \mc{0.814}  & \mc{1.000}   \\
\md{$s_{b4}$}   & \mc{0.541}  & \md{1.000}  & \mc{0.970}  & \mc{1.000}   \\
\hline \multirow{4}{10mm}{core}
&$\betn(\o15)$  &  0.00&  0.00&  0.00&  0.00&  0.00&  0.00&  0.00&  0.00 \\
&$\betn(\f16)$  &  0.16&  0.00& -0.08&  0.00&  0.15&  0.00&  0.14&  0.00 \\
&$-E(\f16)$     &111.42&110.97&115.43&114.64&111.42&111.07&114.29&113.94 \\
&$S_p(\f16)$    & -0.53& -0.99&  0.48& -0.31& -0.53& -0.87& -0.17& -0.52 \\
\hline \multirow{6}{0mm}{[000]1/2}
&$\betn(\ols)$  &  0.00&  0.00&  0.00&  0.00&  0.00&  0.00&  0.00&  0.00 \\
&$\betn(\fls)$  &  0.16&  0.00& -0.09&  0.00&  0.15&  0.00&  0.15&  0.00 \\
&$-E(\fls)$     &125.21&124.75&129.38&128.58&125.72&125.35&128.69&128.36 \\
&$S_p(\fls)$    & -0.27& -0.72&  0.83&  0.02& -0.27& -0.63&  0.12& -0.21 \\
&$B_\la(\fls)$  & 13.79& 13.78& 13.95& 13.94& 14.30& 14.28& 14.40& 14.42 \\
&$\dbl^p(\fls)$	&  0.26&  0.26&  0.35&  0.33&  0.26&  0.24&  0.29&  0.31 \\
\hline \multirow{6}{0mm}{[110]1/2}
&$\betn(\olp)$  & -0.03&  0.00& -0.02&  0.00& -0.01&  0.00& -0.01&  0.00 \\
&$\betn(\flp)$  & -0.13&  0.00& -0.10&  0.00& -0.11&  0.00& -0.10&  0.00 \\
&$-E(\flp)$     &114.59&113.10&118.68&117.54&114.69&112.61&117.59&115.91 \\
&$S_p(\flp)$    &  0.09& -1.35&  1.23&  0.12&  0.09& -1.95&  0.48& -1.18 \\
&$B_\la(\flp)$  &  3.17&  2.13&  3.25&  2.90&  3.27&  1.54&  3.30&  1.97 \\
&$\dbl^p(\flp)$ &  0.62& -0.36&  0.75&  0.43&  0.62& -1.08&  0.65& -0.65 \\
\hline \multirow{6}{0mm}{[101]1/2 [101]3/2}
&$\betn(\olp)$  &  0.04&  0.00&  0.04&  0.00&  0.04&  0.00&  0.03&  0.00 \\
&$\betn(\flp)$  &  0.20&  0.00&  0.16&  0.00&  0.18&  0.00&  0.17&  0.00 \\
&$-E(\flp)$     &115.27&114.03&119.16&118.10&115.27&114.19&118.15&117.13 \\
&$S_p(\flp)$    &  0.64& -0.43&  1.60&  0.67&  0.57& -0.39&  0.94&  0.03 \\
&$B_\la(\flp)$  &  3.85&  3.06&  3.73&  3.46&  3.85&  3.12&  3.86&  3.19 \\
&$\dbl^p(\flp)$ &  1.17&  0.55&  1.12&  0.98&  1.09&  0.48&  1.11&  0.55 \\
\end{tabular}
\end{ruledtabular}
\label{t:16f}
\end{table*}

\begin{figure*}[t]
\hskip-9mm
\includegraphics[width=0.8\linewidth]{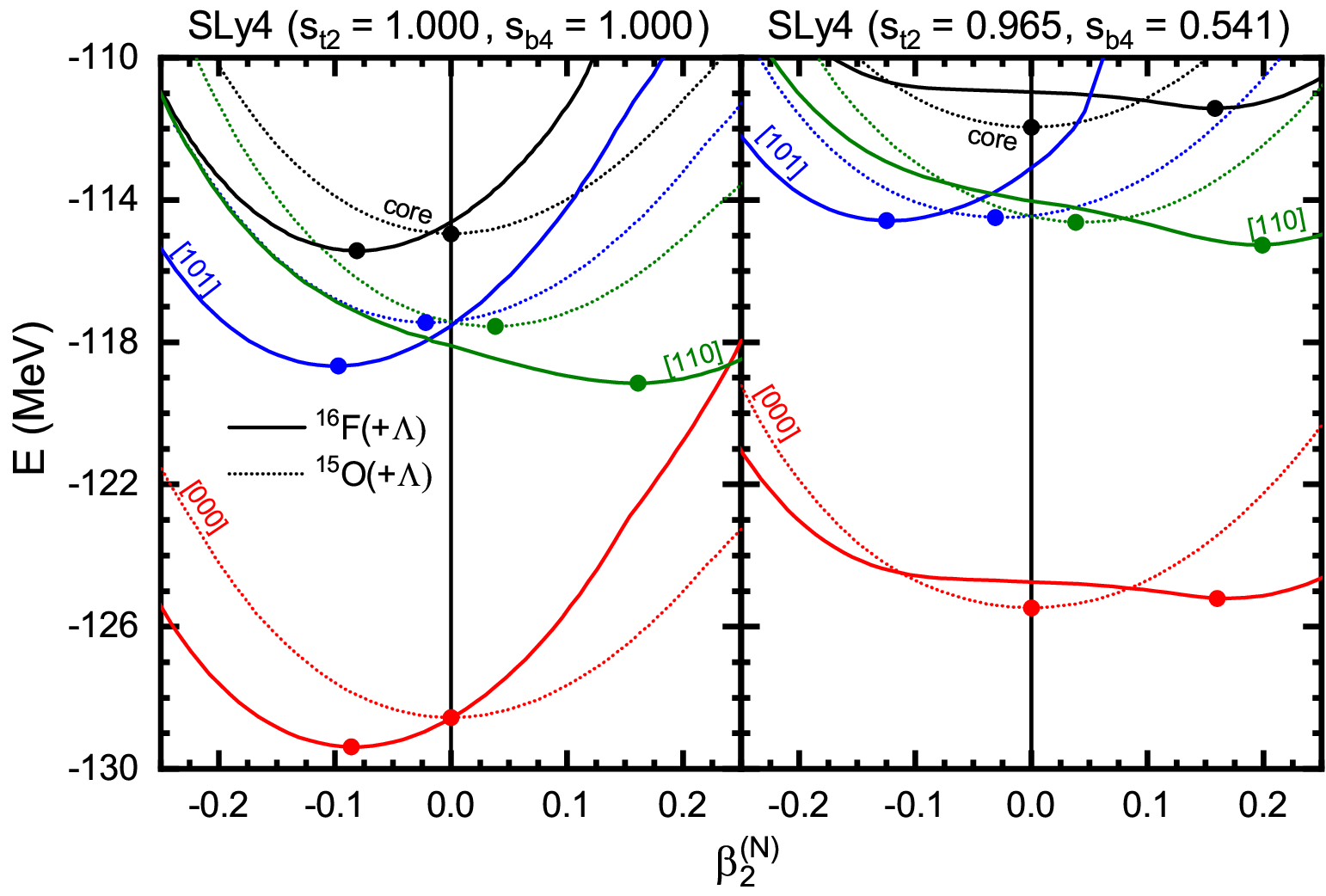}
\vskip-4mm
\caption{
The potential energy surfaces of $^{16}$F and $^{15}$O
(black curves)
and their $s$- and $p$-state $\la$ hypernuclei
(colored curves)
obtained with the original SLy4 force (left)
and the adjusted force (right)
reproducing correctly the energies
$E(^{16}\text{F})=-111.42\mev$ and
$E(^{15}\text{O})=-111.96\mev$.
Markers indicate the energy minima.
\hj{}
}
\label{f:16f}
\end{figure*}

\begin{figure*}[t]
\hskip-9mm
\includegraphics[width=0.8\linewidth]{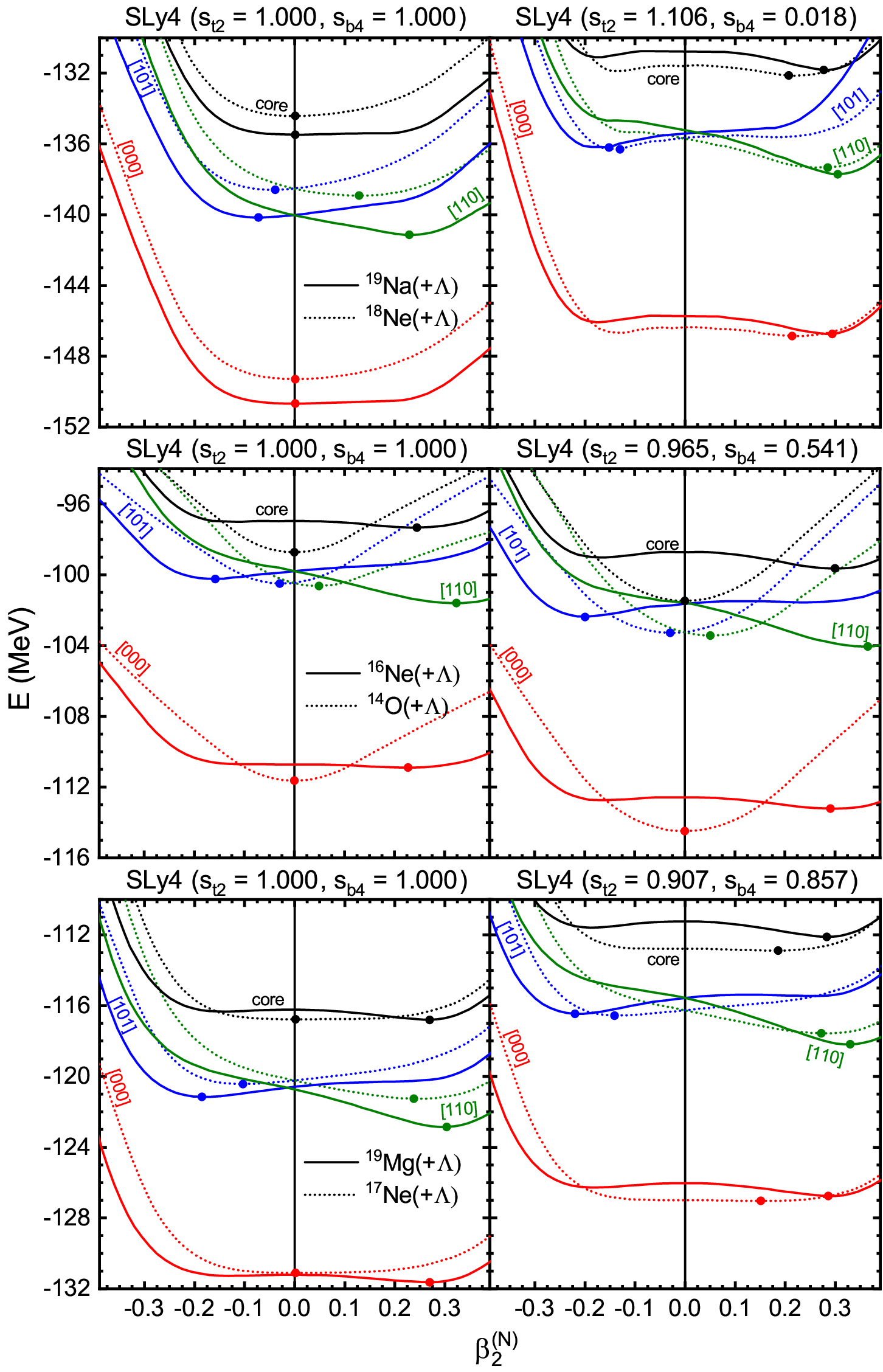}
\vskip-3mm
\caption{
Same as Fig.~\ref{f:16f}, but
(a) for $^{19}$Na and $^{18}$Ne with energies
$E(^{19}\text{Na})=-131.82\mev$ and $E(^{18}\text{Ne})=-132.14\mev$,
(b) for $^{16}$Ne and $^{14}$O with energies
$E(^{16}\text{Ne})=-97.33\mev$ and $E(^{14}\text{O})=-98.73\mev$, and
(c) for $^{19}$Mg and $^{17}$Ne with energies
$E(^{19}\text{Mg})=-112.12\mev$ and $E(^{17}\text{Ne})=-112.88\mev$.
\hj{}
}
\label{f:19na}
\end{figure*}

In the present calculations,
the DSHF Schr\"odinger equation is
solved in cylindrical coordinates $(r,z)$,
under the assumption of axial symmetry of the mean fields.
The optimal quadrupole deformation parameters
\be
\bet^{(q)} = \sqrt{\frac{\pi}{5}}
\frac{\langle 2z^2-r^2 \rangle_q}{\langle z^2+r^2 \rangle_q}
\ee
are calculated by minimizing the binding energy $E(\bet)$.
\OFF{
However, when comparing with experimental deformations derived
from the quadrupole moment $Q_p$, we employ the definition
\be\label{e:beta}
 \beta = \frac{\sqrt{5\pi}}{3} \frac{Q_p}{Z R_0^2} \:
\ee
with $R_0 \equiv 1.2\,A^{1/3}\fm$ \cite{Raman01}.

The rms radius of a hypernucleus is
\be
 R_N = \sqrt{\frac{N}{A+1}R_n^2+\frac{Z}{A+1}R^2_p+\frac{1}{A+1}R^2_\la} \:,
\ee
where $A$ is the total baryon number of the core nucleus,
$Z$ is the proton number, $N$ is the neutron number,
and $R_{n,p,\la}$ is the neutron/proton/$\la$ rms radius, respectively.
}
The single-proton separation energy of a normal nucleus is then
\be
 S_p(^{A}Z) = E(^{A}Z) - E(^{A_-}Z_-) \:,
\ee
with $A_- \equiv A-1$, $Z_- \equiv Z-1$
and similarly for the hypernucleus,
\bal
 S_p(^{A_+}_{\ \la}Z) &= E(^{A_+}_{\ \la}Z) - E(^{A}_{\la}Z_-)
\\
 &= E(^{A} Z) + B_\la(^{A_+}_\la Z) -
    E(^{A_-}Z_-) - B_\la(^{A}_\la Z_-)
\\
 &\equiv S_p(^{A}Z) + \dbl(^{A_+}_\la Z) \:,
\label{e:dbl}
\eal
which relates the proton removal energy of the hypernucleus
to the one of the core nucleus
and the difference $\dbl$ of the $\la$ removal energies
of the two cores involved \cite{Lanskoy22}.
It is the latter that might stabilize the hypernucleus against proton decay
of the otherwise unstable core,
because the $\la$ is stronger bound in the initial nucleus than in the final one,
and this also causes a stronger additional binding of the valence proton.
Equivalent expressions are valid for the two-proton removal energy.

Here the binding energy like $E(^{A}Z)$ is the energy at the lowest point of the
corresponding $E(\bet)$ curve,
which implies that the single-particle (s.p.) energies
for all occupied proton levels are negative (bound).
Vice-versa, however, a bound valence proton s.p.~state does not necessarily
imply a positive separation energy.
Negative $S_p$ corresponds to the experimental situation of a proton emitter
in our model,
and we focus in particular on the isotopes with a small negative
separation energy $-1\mev \lesssim S_p < 0$,
which might be turned to positive in the hypernucleus
by the above mechanism.

\section{Results}
\label{s:res}


We begin in Table~\ref{t:sp} with a list of the single- and double-proton
separation energies of the light-nuclei isotopes just beyond their drip line,
i.e., at least one of $S_p$ or $S_{pp}$ is negative.
If an added $\la$ hyperon is able to stabilize a nuclear core sufficiently
to render both $S_p$ and $S_{pp}$ positive,
that hypernucleus becomes stable.
This will only occur if the problematic separation energy is not too large
(negative),
typically $S\gtrsim-1\mev$ for medium nuclei.
According to the table, good candidates are
$^{7}$B, $^{16}$F, $^{19}$Na, $^{30}$Cl
(critical, i.e., more negative, $S_p$)
and
$^{6}$Be, $^{8}$C, $^{12}$O, $^{16}$Ne, $^{19}$Mg
(critical $S_{pp}$).
In fact those have also been identified in \cite{Lanskoy22,Kornilova23}
as possible candidates.

The only experimentally known hypernucleus exhibiting this phenomenon
is the lightest one in the table, $^7_\la$Be,
with a $\la$ removal energy $B_\la=5.16\pm0.09\mev$ \cite{Davis86,HyperChart}
and a two-proton removal energy $S_{pp}=+0.69\pm0.12\mev$
to ${^5_\la\text{He}}$ with $B_\la=3.10\pm0.03\mev$,   
compared to $S_{pp}({^6\text{Be}}\ra{^4\text{He}}) = -1.37\mev$ \cite{NuDat}
for the core nucleus.
Therefore we now investigate the heavier candidates in our theoretical approach,
beginning with the most promising case of the $^{16}$F core
and its critical single-proton decay to $^{15}$O.

A problem of the SHF approach with a common $NN$ Skyrme force is that
the experimental binding energies of any nuclei are never exactly reproduced,
but in our case accuracies of order $0.1\mev$ are required for reliable
predictions of the removal energies.
For example, the experimental energies are
$E(\f16)=-111.42\mev$,
$E(\o15)=-111.95\mev$,
and therefore $S_p=-0.53\mev$,
whereas with the SLy4 force (and spherical calculation) the predictions are
$E(\f16)=-114.64\mev$,
$E(\o15)=-114.95\mev$,
and $S_p=-0.31\mev$.
For other nuclei, the differences may be even larger and of different sign,
i.e., predicting a wrong drip line nucleus.

We compare in the following two methods to alleviate this problem:
(a) As in \cite{Lanskoy22},
we compute the quantity $\dbl$, Eq.~(\ref{e:dbl}),
using the original unmodified Skyrme force.
Added to the experimental $S_p$ value of the core,
it provides an estimate of $S_p$ for the hypernucleus.
(b) We adjust two $NN$ Skyrme force parameters such that the energies of
both core nuclei, e.g., $\f16$ and $\o15$,
(and therefore also $S_p$)
are reproduced exactly by the DSHF calculation.
In practice, we have chosen to adjust the surface parameter $t_2$
and the s.o.~parameter $b_4$ of the (SLy4 or SkI4) force,
introducing scaling parameters $s_{t2}$ and $s_{b4}$.
The motivation for this choice is that naively the
extended proton valence orbits
are most sensitive to these parameters of the force,
but other choices are in principal possible.
The comparison of the SLy4 and SkI4 results will also reveal
how reliable this method is.

Another very important issue is the effect of core deformation,
and we compare in the following results obtained in
deformed and spherical calculations,
for both forces and both methods of computing $S_p$.

The results for $\f16$ are shown in Fig.~\ref{f:16f} and Table~\ref{t:16f}.
The figure shows the potential energy surfaces (PESs)
of $\f16$ (solid curves)
and $\o15$ (dashed curves)
and their hypernuclei,
obtained with the original (left panel)
and adjusted (right panel) SLy4 force.
In this case scaling parameters $s_{t2}=0.956$ and $s_{b4}=0.541$
are required in order to fit exactly the binding energies of both nuclei.
The core nuclei (black curves)
are compared with $\la~s$-state hypernuclei (red curves)
and $\la~p$-state hypernuclei in the two substates
[101] (blue curves) and [110] (green curves)
with oblate and prolate deformation, respectively.

\begin{table}[t]
\caption{
Predicted critical proton separation energies $S_p$ or $S_{pp}$
with the adjusted SLy4 force
for the drip line $\la$ hypernuclei candidates
with the $\la$ occupying different orbitals.
The deformations $\bet$ of the core nuclei and the $\la$ removal energies
$B_\la$ of the hypernuclei are also given.
Bold face indicates stable hypernuclei.
}
\renewcommand\arraystretch{1.3}
\setlength{\tabcolsep}{1.5pt}
\def\mc#1{\multicolumn{1}{c}{$#1$}}
\def\md#1{\multicolumn{2}{c}{#1}}
\begin{ruledtabular}
\begin{tabular}{ l d rr rr rr rr }
   & \mc{S}   & \md{core} &\md{$s[000]$}&\md{$p[101]$}&\md{$p[110]$} \\
   & \mc{S}   & \mc{S}&\mc{\bet}&\mc{S}&\mc{\bl}&\mc{S}&\mc{\bl}&\mc{S}&\mc{\bl} \\
\hline
$^{9}_{\la}$C    &S_{pp}&-2.11 & 0.00 &-0.10& 7.6 &       &     &       &    \\
$^{12}_{\ \la}$N  & S_p &-1.38 & 0.00 &-0.91&10.5 &  -0.78&-0.3 &  -1.07& 0.1\\
$^{13}_{\ \la}$O &S_{pp}&-1.74 & 0.00 &-0.80&11.0 &  -0.53& 0.3 &  -0.98& 0.5\\
$^{17}_{\ \la}$F  & S_p &-0.53 & 0.16 &-0.27&13.8 &\bf0.09& 3.2 &\bf0.64& 3.9\\
$^{17}_{\ \la}$Ne&S_{pp}&-1.40 & 0.25 &-0.74&13.6 &  -0.25& 2.9 &\bf0.97& 4.3\\
$^{20}_{\ \la}$Na & S_p &-0.32 & 0.28 &-0.12&14.9 &  -0.11& 4.4 &\bf0.38& 5.9\\
$^{20}_{\ \la}$Mg&S_{pp}&-0.77 & 0.28 &-0.28&14.7 &  -0.10& 4.3 &\bf0.63& 6.1\\
$^{31}_{\ \la}$Cl & S_p &-0.48 & 0.12 &-0.38&18.3 &  -0.11& 8.9 &  -0.02& 9.8\\
\end{tabular}
\end{ruledtabular}
\label{t:lsp}
\end{table}

It can be seen that the adjusted force reproduces correctly
the energies of both $\f16$ and $\o15$ in their deformation minima,
whereas the original force overbinds both nuclei by about $3\mev$.
In the hypernucleus $^{17}_\la$F with the adjusted force,
an $s$-state $\la$ is bound by $\bl=13.8\mev$
and a $p$-state $\la$ by
$\bl=3.2\mev$ ([101]) or $\bl=3.9\mev$ ([110]).
In the latter case both nuclear core and hyperon orbital are prolately
deformed with $\bet\approx0.2$,
which causes an important gain in binding energy
\cite{Guo21,Xue22,Liu23,Xue23,Xue24}.

This is also decisive for the effect of the hyperon
on the proton separation energy,
which can better be seen by the precise values listed in Table~\ref{t:16f}.
Namely, the proton separation energies
for $s$[000], $p$[101], $p$[110] hypernuclei
are $S_p=-0.27,0.09,0.64\mev$, respectively,
compared to $-0.53\mev$ of the core nucleus.
Thus the coinciding deformation characteristics of core and $p$[110] hyperon
indeed bind also the valence proton sufficiently strong
to make the hypernucleus stable.
This is not the case for a hyperon in the spherical $s$[000] state,
or for the spherical calculation (4th column of Table~\ref{t:16f}),
where all hypernuclei remain unbound.
Therefore an inclusion of deformation in the theoretical formalism
is essential for our results.

Regarding the approximate calculation of the hypernucleus $S_p$
by adding $\dbl$ to the cores's experimental value,
the $\dbl$ values are similar for the deformed calculations
with adjusted and original SLy4 and SkI4 forces (columns 3, 5, 7, 9),
as long as the predicted deformations are also similar.
The $\dbl$ of the spherical calculations (columns 4, 6, 8, 10) are also similar
(to those of the deformed calculations)
for the compatible $s$[000] state
(our results agree with those of \cite{Lanskoy22}),
but not for the $p$-state hyperons.
Thus this approximation can only be used with confidence for the $\la~s$ states.

The previous discussion pertains also to the other (hyper)nuclei we investigated
and the results are summarized in Table~\ref{t:lsp},
where we list the values of the relevant proton separation energies $S$
and also the removal energies $\bl$ of the various $\la$ orbitals.
The results show that the hypernuclei
$\flp$, $^{20}_{\la p}$Na, $^{17}_{\la p}$Ne, $^{20}_{\la p}$Mg
become stable,
thus extending the proton drip point of their core nuclei.
There are, however, no stable $s$-state hypernuclei
(as in \cite{Lanskoy22}),
hence the deformation effect is essential for the strong binding
of the valence protons
and the $\la~p$ states (several MeV) in those hypernuclei.
For completeness, the PESs for the stable nuclei
$^{20}_{\la p}$Na, $^{17}_{\la p}$Ne, $^{20}_{\la p}$Mg
are shown in Fig.~\ref{f:19na}. 
They are all strongly prolate deformed ($\bet\approx0.3$),
and thus the binding of the $\la~p$ [110] state
is enhanced by the same mechanism.
We refer to the analogous discussion of $\flp$ in Fig.~\ref{f:16f} above.
The remaining hypernuclei result unstable in all types of calculations
we performed.

Some comments are in order for $^{9}_{\la}$C:
In \cite{Lanskoy22} this hypernucleus was claimed to be stable with
$\dbl\approx2.5\mev$,
added to $S_{pp}=2.11\mev$ of $^{8}$C.
However, in this particular case $\dbl$ was calculated as the difference
of the theoretical $\bl$ for $^{9}_{\la}$C
and the experimental value for $^{7}_{\la}$Be,
breaking the consistency of the procedure
(the theoretical value might be too large;
theoretical errors are not compensating;
see further discussion in \cite{Lanskoy22}).
Using theoretical values for both hypernuclei we obtain instead
$\dbl\approx2.01\mev$,
and thus a marginable unstable hypernucleus.
This delicate case should be studied in more detail in the future,
perhaps within a more suitable few-body model.

\section{Summary}
\label{s:conc}

We investigated (hyper)nuclei at the proton drip line
in the framework of the deformed SHF mean-field model
with a combination of the $NN$ interaction SLy4 or SkI4 and
the $\la N$ interaction SLL4.
The binding energies of the relevant core nuclei were fitted exactly
by adjusting the parameters $t_2$ and $b_4$ of the original Skyrme $NN$ forces.
Then accurate values of the proton separation energies were obtained
for the core nucleus and for the hypernucleus
with a $\la$ in either $s$ or $p$ states.
The results are nearly independent of the $NN$ force.
We find that several deformed-core nuclei might become stabilized by the addition
of a $\la$ in a $p$ substate with the same deformation as the core,
whereas the effect of a $s$-state $\la$ is not sufficiently strong.
Our results affirm that
$\flp$, $^{17}_{\la p}$Ne, $^{20}_{\la p}$Na, $^{20}_{\la p}$Mg
become stable nuclei,
thus extending the drip point of their cores.

We compared the results with an approximate correction
to the actual proton separation energy given by the
difference of the $\la$ removal energies of the involved hypernuclei,
employing the unmodified $NN$ forces.
This approximation is only reliable for the $s$-state $\la$ hypernuclei,
which are, however, not stable.

Thus deformed calculations allowing the gain of binding energy
of the $p$-state $\la$ are essential for our predictions.
Our results have to be confirmed by more accurate beyond-mean-field calculations,
in particular regarding the theoretical modeling
of deformation and weakly-bound states.
Future experimental data will be very valuable to improve the theoretical models.

\section*{Acknowledgments}

This work was supported by the National Natural Science Foundation of China
under Grant Nos.~12175071, 12205103.

\newcommand{\araa}{Annu. Rev. Astron. Astrophys.}
\newcommand{\cp}{Contemp. Phys.}
\newcommand{\cpc}{Chin. Phys. C}
\newcommand{\epja}{EPJA}
\newcommand{\jpg}{J. Phys. G}
\newcommand{\npa}{Nucl. Phys. A}
\newcommand{\nphysa}{Nucl. Phys. A}
\newcommand{\physrep}{Phys. Rep.}
\newcommand{\plb}{Phys. Lett. B}
\newcommand{\ppnp}{Prog. Part. Nucl. Phys.}
\newcommand{\ptp}{Prog. Theor. Phys.}
\newcommand{\ptep}{Prog. Theor. Exp. Phys.}
\newcommand{\ppnl}{Phys. Part. Nucl. Lett.}
\bibliographystyle{apsrev4-1}
\bibliography{pdrip}

\end{document}